\documentstyle[11pt,a4]{article}
\newcommand{\QED}{{\hfill$\Box$}}
\newcommand{\du}{{$\downarrow$}}

\begin{document}
\bibliographystyle{plain}
\title{Uses of Randomness in Computation%
\thanks{Copyright \copyright\ 1994, R.~P.~Brent.
Text of a talk given at University of NSW (Theory Day),
Griffith University and ANU, April--May 1994.
}}
\author{Richard P.\ Brent%
\thanks{E-mail address: {\tt rpb@cslab.anu.edu.au}
	\hspace*{\fill} rpb147 typeset using \LaTeX}\\
	Computer Sciences Laboratory\\
	Australian National University\\
	Canberra, ACT 0200, Australia\\[15pt]
	Report TR-CS-94-06\\
	21 June 1994}
\date{}
\maketitle
\thispagestyle{empty}			%
\vspace{-0.8cm}
\begin{abstract}

Random number generators are widely used in practical algorithms.
Examples include simulation, number theory (primality testing and integer
factorization), fault tolerance, routing, cryptography, optimization by
simulated annealing, and perfect hashing.
\smallskip

Complexity theory usually considers the worst-case behaviour of
deterministic algorithms, but it can also consider average-case behaviour
if it is assumed that the input data is drawn randomly from a given
distribution. Rabin popularised the idea of ``probabilistic'' algorithms,
where randomness is incorporated into the algorithm instead of being assumed
in the input data. Yao showed that there is a close connection between the
complexity of probabilistic algorithms and the average-case complexity of
deterministic algorithms.
\smallskip

We give examples of the uses of randomness in computation,
discuss the contributions of Rabin, Yao and others, and mention some
open questions.

\smallskip

{\em 1991 Mathematics Subject Classification.}
Primary
68-01, 68Q25;
Secondary
05C80, 11A51, 11K45, 34F05, 65C10, 68P10, 68Q05, 68Q10, 68Q15
\smallskip

{\em Key words and phrases.}
Galileo,
integer factorisation,
Las Vegas algorithm,
Library of Congress on Mars,
minimal perfect hashing,
Monte Carlo algorithm,
perfect hashing,
perfect party problem,
permutation routing,
primality testing,
probabilistic algorithm,
Ramsey number,
random algorithm,
randomisation,
randomness,
RP.
\end{abstract}

\pagebreak[3]
\section{Checking out Galileo}

The Galileo spacecraft is somewhere near Jupiter, but its main radio
antenna is not working, so communication with it is very slow.
Suppose we want to check that a critical program in
Galileo's memory is correct, and has not been corrupted by a passing
cosmic ray. How can we do this without transmitting the whole program
to or from Galileo~?
\medskip

Here is one way.
The program we want to check (say $N_1$) and the correct
program on Earth (say $N_2$) can be regarded as multiple-precision integers.
Choose a random prime number $p$ in the interval $(10^9, 2\times 10^9)$.
Transmit $p$ to Galileo and ask it to compute
\[ r_1 \leftarrow N_1 \bmod p \]
and send it back to Earth.
Only a few bits (no more than 64 for $p$ and $r_1$)
need be transmitted between Earth and Galileo, so we can afford to use good
error correction/detection.

On Earth we compute
$r_2 \leftarrow N_2 \bmod p$, and check if $r_1 = r_2$.
There are two possibilities:

\begin{itemize}

\item $r_1 \ne r_2$. We conclude that $N_1 \ne N_2$. Galileo's program has
been corrupted~!  If there are only a small number of errors, they can be
localised by binary search using
$O(\log\log N_1)$ %
small messages.

\item $r_1 = r_2$. We conclude that Galileo's program is {\em probably}
correct. More precisely, if Galileo's program is {\em not} correct there
is only a probability of less than $10^{-9}$ that $r_1 = r_2$, i.e. that we
have a ``false positive''.
If this probability is too large for the quality-assurance team to
accept, just repeat the process (say) ten times with different random
primes $p_1, p_2, \ldots, p_{10}$.
If $N_1 \ne N_2$, there is a probability of less than
\[ 10^{-90} \]
that we get $r_1 = r_2$ ten times in a row. This should be good enough.

\end{itemize}

\noindent{The problem and its solution were communicated to me by
Michael Rabin, who called it the ``Library of Congress on Mars'' problem.}

\medskip
\noindent{\em The Structure}
\medskip

Our procedure has the following form. We ask a
question with a yes/no answer. The precise question depends on a random
number. If the answer is ``no'', we can assume that it is correct.
If the answer is ``yes'', there is a small probability of error,
but we can reduce this probability to a negligible level by
repeating the procedure a few times with {\em independent} random numbers.
\medskip

We call such a procedure a {\em probabilistic} algorithm;
other common names are {\em randomised} algorithm
and {\em Monte Carlo} algorithm. 	%

\medskip
\noindent{\em Disclaimer}
\medskip

It would be much better to build error correcting hardware into Galileo,
and not depend on checking from Earth.

\section{Testing Primality}

Here is another example\footnote{Due to M.~O.~Rabin~\cite{Rab76},
with improvements by G.~L.~Miller. See Knuth~\cite{Knuth-2}.}
with the same structure. We want an algorithm
to determine if a given odd positive integer $n$ is prime.
Write $n$ as ${2^k}q + 1$, where $q$ is odd and $k > 0$.
\medskip

\noindent{\em Algorithm~P}

\begin{enumerate}

\item Choose a random integer $x$ in $(1, n)$.

\item Compute $y = x^q \bmod n$. This can be done with $O(\log q)$ operations
mod~$n$, using the binary representation of $q$.

\item If $y = 1$ then return ``yes''.

\item For $j = 1, 2, \ldots, k$ do
	\begin{description}
	\item[]	if $y = n-1$ then return ``yes''
	\item[]	else if $y = 1$ then return ``no''
	\item[] else $y \leftarrow y^2 \bmod n$. 
	\end{description}

\item Return ``no''.

\end{enumerate}

\noindent{\em Fermat's Little Theorem}
\medskip

To understand the mathematical basis for Algorithm~P, recall Fermat's
little Theorem:\\
if $n$ is prime and $0 < x < n$, then
\[ x^{n-1} = 1 \bmod n. \]
Thus, if $x^{n-1} \ne 1 \bmod n$, we can definitely say that $n$ is
composite.
\medskip

Unfortunately, the converse of Fermat's little theorem is
false: if $x^{n-1} = 1 \bmod n$ we can not be sure that $n$ is prime.
There are examples (called {\em Carmichael numbers}) of composite $n$
for which $x^{n-1}$ is always $1 \bmod n$ when ${\rm{GCD}}(x,n) = 1$.
The smallest example is
\[ 561 = 3 \cdot 11 \cdot 17 \]
Another example is\footnote{Hardy's taxi number~\cite{Hardy-Ramanujan},
$1729 = 12^3 + 1^3 = 10^3 + 9^3$.}
\[ n = 1729 = 7 \cdot 13 \cdot 19 \]

\noindent{\em An Extension}
\medskip

A slight extension of Fermat's little Theorem is useful,
because its converse is {\em usually} true.
\medskip

If $n = {2^k}q + 1$
is an odd prime, then either $x^q = 1 \bmod n$, or the sequence
\[ \left(x^{{2^j}q} \bmod n\right)_{j=0,1,\ldots,k} \]
ends with $1$, and the value just preceding the first appearance of $1$
must be $n-1$.
\medskip

{\em Proof:} If $y^2 = 1 \bmod n$ then $n | (y-1)(y+1)$.
Since $n$ is prime, $n | (y-1)$ or $n | (y+1)$.
Thus $y = \pm 1 \bmod n$. \QED
\medskip

The extension gives a {\em necessary} (but not sufficient) condition
for primality of $n$. Algorithm~P just checks if this condition is
satisfied for a random choice of $x$, and returns ``yes'' if it is.

\medskip
\noindent{\em Reliability of Algorithm~P}
\medskip

Algorithm~P can not give false negatives (unless we make an arithmetic
mistake), but it can give false positives
(i.e. ``yes'' when $n$ is composite).
However, the probability of a false positive is less than $1/4$.
(Usually much less~-- see Knuth~\cite{Knuth-2}, ex.~4.5.4.22.)
Thus, if we repeat the algorithm $10$ times there is less than 1 in $10^6$
chance of a false positive, and if we repeat $100$ times the results should
satisfy anyone but a pure mathematician.
\medskip

Algorithm P works fine even if the input is a Carmichael number.

\medskip

\noindent{\em Use of Randomness}
\medskip

Note that in both our examples randomness was introduced into the algorithm.

{\bf We did not make any assumption about the distribution of inputs.}

\medskip
\noindent{\em Summary of Algorithm~P}
\medskip

Given any $\varepsilon > 0$, we can check primality of a
number $n$ in
\[ O((\log n)^3 \log(1/\varepsilon)) \]
bit-operations\footnote{We can factor $n$ deterministically in
$O(\log n)$ %
{\em arithmetic} operations~\cite{Sha79},
but this result is useless because the
operations are on numbers as large as $2^n$.
Thus, it is more realistic to consider bit-operations.},
provided we are willing to accept a probability of error
of at most~$\varepsilon$.
\medskip

By way of comparison, the best known {\em deterministic} algorithm
takes					%
\[ O((\log n)^{c\log\log\log n}) \]	%
bit-operations,				%
and is much more complicated.
If we assume the {\em Generalised Riemann Hypothesis}, the exponent
can be reduced to $5$.			%
(But who believes in GRH with as much certainty as Algorithm~P
gives us~?)

\section{Error-Free Algorithms}

The probabilistic algorithms considered so far ({\em Monte Carlo} algorithms)
can give the wrong
answer with a small probability. There is another class of
probabilistic algorithms
({\em Las Vegas} algorithms)
for which the answer is always correct;
only the runtime is random\footnote{In practical cases the
expected runtime is finite. It is possible that the
algorithm does not terminate, but with probability zero.}.
An interesting example is
H.~W.~Lenstra's {\em elliptic curve method} (ECM)~\cite{Len87a}
for integer factorisation.
To avoid trivial cases, suppose we want to find a
prime factor $p>3$ of an odd composite integer~$N$.
\medskip

To motivate ECM, consider an earlier algorithm, Pollard's ``$p-1$'' method.
This works if
$p-1$ is ``smooth'', i.e.~has only small prime factors. $p-1$ is
important because it is the order of the multiplicative group $G$ of
the field $F_p$. The problem is that $G$ is fixed.

\medskip
\noindent{\em Lenstra's Idea}
\medskip

Lenstra had the idea of using a group $G(a,b)$ which depends on
parameters $(a,b)$. By randomly selecting $a$ and $b$, we get a large set
of different groups, and some of these should have smooth order.
\medskip

The group $G(a,b)$ is the group of points on the {\em elliptic curve}
\[ y^2 = x^3 + ax + b \bmod p, \]		%
and by a famous theorem\footnote{The ``Riemann hypothesis for finite fields''.
$G(a,b)$ is known as the ``Mordell-Weil'' group. The result on its order
follows from a theorem of Hasse~(1934),	%
later generalised by A.~Weil and Deligne (see~\cite{Len90}).}
the order of $G(a,b)$ is an integer in the interval
\[ (p -1 - 2\sqrt{p},\;\; p -1 + 2\sqrt{p}) \]
The distribution in this interval is not uniform, but it is ``close enough''
to uniform for our purposes.

\medskip
\noindent{\em Runtime of ECM}
\medskip

Under plausible assumptions ECM has expected run time
\[ T = O\left(\exp (\sqrt{c\log p \log\log p}) (\log N)^2\right),\]
where $c \simeq 2$.
\medskip

Note that $T$ depends mainly on the
size of $p$, the factor found, and not very strongly on $N$.
In practice the run time is close to an exponentially distributed random
variable with mean and variance about $T$.

\medskip
\noindent{\em ECM Example}
\medskip

ECM is the best known algorithm for finding
moderately large factors of very large numbers.
\medskip

Consider the 617-decimal digit Fermat number
$F_{11} = 2^{2^{11}} + 1$. Its factorisation is:
\begin{eqnarray*}
F_{11}	&=& 319489 \cdot 974849 \cdot
	167988556341760475137 \cdot
	3560841906445833920513 \cdot p_{564},
\end{eqnarray*}
where $p_{564}$ is a 564-decimal digit prime.
\medskip

In 1989 I found the 21-digit and 22-digit prime factors using ECM.
The factorisation required about 360 million multiplications mod $N$,
which took less than 2 hours on a Fujitsu VP~100 vector processor.

\section{Minimal Perfect Hashing}

{\em Hashing} is a common technique used to map words into a small set of
integers (which may then be used as indices to address a table). Thus, the
computation $r_1 \leftarrow N_1 \bmod p$ used in our ``Galileo'' example
can be considered as a hash function.
\medskip

Formally, consider a set
\[ W = \{w_0, w_1, \ldots, w_{m-1}\} \]
of $m$ words $w_j$, each of which is a finite string
of symbols over a finite alphabet $\Sigma$. A {\em hash function} is a
function
\[ h : W \to I, \]
where $I = \{0, 1, \ldots, k-1\}$ and $k$ is a fixed
integer (the table size).

\medskip
\noindent{\em Collisions}
\medskip

A {\em collision} occurs if two words $w_1$ and $w_2$ map to the same address,
i.e.~if $h(w_1) = h(w_2)$. There are various techniques for handling
collisions~\cite{Knuth-3}.
However, these complicate the algorithms and introduce inefficiencies.
In applications where $W$ is fixed (e.g.~the reserved words in a compiler),
it is worth trying to avoid collisions.

\medskip
\noindent{\em Perfection}
\medskip

If there are no collisions, the hash function
is called {\em perfect}.

\medskip
\noindent{\em Minimal Perfection}
\medskip

For a perfect hash function, we must have $k \ge m$. If $k = m$ the
hash function is {\em minimal}.

\medskip
\noindent{\em Problem}
\medskip

Given a set $W$, how can we compute a minimal perfect hash function~?

\medskip
\noindent{\em The CHM Algorithm}
\medskip

Czech, Havas and Majewski (CHM)~\cite{Cze92}
give a probabilistic algorithm
which runs in expected time $O(m)$ (ignoring the effect of finite word-length).
Their algorithm uses some properties of {\em random graphs}.
\medskip

Take $n = 3m$, 
and let
\[ V = \{1, 2, \ldots, n\}. \]
CHM take two independent pseudo-random functions\footnote{How can this be
done~? This is a
theoretical weak point of the algorithm, but in practice the solution
given in~\cite{Cze92} is satisfactory.}
\[ f_1 : W \to V,\;\; f_2 : W \to V, \]
and let
\[ E = \{(f_1(w), f_2(w))\;|\; w \in W\}. \]
\medskip

We can think of $G = (V,E)$ as a random graph with $n$ vertices $V$
and (at most) $m$ edges $E$.

\medskip
\noindent{\em Acyclicity}
\medskip

If $G$ has less than $m$ edges or $G$ has
cycles, CHM reject the choice of $f_1, f_2$ and try again. Eventually
they get a graph $G$ with $m$ edges and no cycles. Because $n = 3m$,
the expected number of trials is a constant (about
$\sqrt{3}$, or more generally $\sqrt{{n}\over{n-2m}}$,
for large $m$ and $n > 2m$).

\pagebreak[4]
\noindent{\em The Perfect Hash Function}
\medskip

Once an acceptable $G$ has been found, it is easy to compute
(and store in a table) a function
\[ g : V \to {0, 1, \ldots, m-1} \]
such that
\[ h(w) = g(f_1(w)) + g(f_2(w)) \bmod m \] %
is the desired minimal perfect hash function. We can even get
\[ h(w_j) = j \]
for $j = 0, 1, \ldots, m-1$.
All this requires is a depth-first search of $G$.

\medskip
\noindent{\em Implementation}
\medskip

CHM report that on a Sun SPARCstation~2 they can generate
a minimal perfect hash function for a set of $m = 2^{19}$ words in
$33$ seconds. Earlier algorithms required time which (at least in the
worst case) was an exponentially increasing function of $m$, so could
only handle very small $m$.

\section{Permutation Routing}
\medskip

A network {\cal G} is a connected, undirected graph with $N$ vertices
$0, 1, \ldots, N-1$.
\medskip

The {permutation routing} problem on {\cal G} is:
given a permutation $\pi$ of the vertices, and a message (called
a {\em packet}) on each vertex, route packet $j$
from vertex $j$ to vertex $\pi(j)$.
It is assumed that at most one packet can traverse each edge in unit
time, and that we want to minimise the time for the routing.
\medskip

In practice we only want to consider {\em oblivious} algorithms,
where the route taken by packet $j$ depends only on $(j, \pi(j))$.
\medskip

For simplicity, assume that the {\cal G} is a $d$-dimensional
hypercube, so $N = 2^d$.
Similar results apply to other networks.

\medskip
\noindent{\em Example: Leading Bit Routing}
\medskip

A simple algorithm for routing packets on a hypercube chooses
which edge to send a packet along by comparing the current address
and the destination address and finding the highest order bit position
in which these addresses differ. %

For example, consider the bit-reversal permutation
$01001001 \rightarrow 10010010$. Each ``\du'' corresponds to
traversal of an edge in the hypercube.
\medskip

\centerline{			%
\begin{tabular}{ccccccccc}
	&0&1&0&0&1&0&0&1\\
	&\du&&&&&&&\\
	&1&1&0&0&1&0&0&1\\
	&&\du&&&&&&\\
	&1&0&0&0&1&0&0&1\\
	&&&&\du&&&&\\
	&1&0&0&1&1&0&0&1\\
	&&&&&\du&&&\\
	&1&0&0&1&0&0&0&1\\
	&&&&&&&\du&\\
	&1&0&0&1&0&0&1&1\\
	&&&&&&&&\du\\
	&1&0&0&1&0&0&1&0\\
\end{tabular}
}

\pagebreak[4]
\noindent{\em Borodin and Hopcroft's bound}
\medskip

The following result~\cite{Bor85} says that there are no ``uniformly good''
deterministic algorithms for oblivious permutation routing:
\medskip

\noindent{\em Theorem:} For any deterministic, oblivious permutation routing
algorithm, there is a permutation $\pi$ for which the routing takes
$\Omega(\sqrt{N/d^3})$ steps.
\medskip

\noindent{\em Example:} For the leading-bit routing algorithm,
take $\pi$ to be the bit-reversal permutation,~i.e.
$$\pi(b_0b_1\ldots b_{d-1}) = b_{d-1}\ldots b_1b_0\;. $$
Suppose $d$ is even.
Then at least $2^{d/2}$ packets are routed through vertex~$0$.
To prove this, consider the routing of
$$xx \ldots xx00 \ldots 00\;,$$
where there are at least $d/2$ trailing zeros.

\medskip
\noindent{\em Valiant and Brebner's algorithm}
\medskip

We can do much better with a probabilistic algorithm.
Valiant suggested:
\begin{enumerate}
\item Choose a random mapping $\sigma$ (not necessarily a permutation).
\item Route message $j$ from vertex $j$ to vertex $\sigma(j)$
	using the leading bit algorithm
	(for $0 \le j < N$).
\item Route message $j$ from vertex $\sigma(j)$ to vertex $\pi(j)$.
\end{enumerate}

This seems crazy\footnote{I do not know of any manufacturer who has
been persuaded to implement it.
Probably it would be hard to sell.}, but it works~!
Valiant and Brebner~\cite{Val81}
prove:
\medskip

\noindent{\em Theorem:} With probability greater than $1 - 1/N$, every
packet reaches its destination in at most $14d$ steps.
\medskip

\noindent{\em Corollary:} The expected number of steps to route all packets
is less than $15d$.

\section{Pseudo-deterministic Algorithms}

Some probabilistic algorithms use many independent random numbers,
and because of the ``law of large numbers'' their performance is
very predictable. One example is the {\em multiple-polynomial quadratic
sieve} (MPQS) algorithm for integer factorisation.
\medskip

Suppose we want to factor a large composite number $N$ (not a perfect power).
The key idea of MPQS is to generate a sufficiently large number of congruences of the
form
\[ y^2 = p_1^{\alpha_1}\cdots p_k^{\alpha_k} \bmod N, \]
where $p_1, \ldots, p_k$ are small primes in a precomputed
``factor base'', and $y$ is close to $\sqrt{N}$. Many $y$ are tried,
and the ``successful'' ones are found efficiently by a sieving process.
\medskip

Making some plausible assumptions, the expected run time of MPQS is
\[ T = O(\exp (\sqrt{c\log N \log\log N})),	\]
	where $c \simeq 1.$
In practice, this estimate is good and the variance is small. %

\pagebreak[4]
\noindent{\em MPQS Example}
\medskip

MPQS is currently the best general-purpose algorithm for factoring moderately
large numbers $N$ whose factors are in the range $N^{1/3}$ to $N^{1/2}$.
For example, A.~K.~Lenstra and M.~S.~Manasse recently found	%
\begin{eqnarray*}
3^{329}+1 &=& 2^2\cdot 547\cdot 16921\cdot 256057\cdot
	36913801\cdot 177140839\cdot
	1534179947851\cdot p_{50} \cdot p_{67}\;,
\end{eqnarray*}
where the penultimate factor $p_{50}$ is a 50-digit prime
$$ 24677078822840014266652779036768062918372697435241, $$
and the largest factor $p_{67}$ is a 67-digit prime.
\medskip

The computation used a network of
workstations for ``sieving'', then a super-computer	%
for the solution of a very large linear system.
\medskip

A ``random'' 129-digit number (RSA129) has just been factored in a similar
way to win a \$100 prize offered by Rivest, Shamir and Adleman in 1977.

\section{Complexity Theory of Probabilistic Algorithms}

Do probabilistic algorithms have an advantage over deterministic
algorithms~? If we allow a small probability of error, the answer is
{\bf yes}, as we saw for the Galileo example. If no error is allowed,
the answer is (probably) {\bf no}.
\medskip

A.~C.~Yao considered probabilistic algorithms (modelled as decision trees)
for testing properties $P$ of undirected graphs (given by their adjacency
matrices) on $n$ vertices.
He also considered deterministic algorithms which assume a given distribution
of inputs (i.e.~a distribution over the set of graphs with $n$ vertices).

\medskip
\noindent{\em Definitions}
\medskip

Yao defines
\medskip

{\em randomized complexity} $F_R(P)$ as an
\medskip

\noindent\hspace*{6em} {\bf infimum} (over all possible algorithms) of a\\
\hspace*{10em}   {\bf maximum} (over all graphs
 		       with $n$ vertices) of the\\
\hspace*{14em}     {\bf expected} runtime.

\medskip
and
\medskip

{\em distributional complexity} $F_D(P)$ as a
\medskip

\noindent\hspace*{6em} {\bf supremum} (over input distributions) of a\\
\hspace*{10em}   {\bf minimum} (over all possible
		       deterministic algorithms) of the\\
\hspace*{14em}       {\bf average} runtime.
\medskip

Informally,
$F_R(P)$ is how long the best probabilistic algorithm takes for testing $P$;
and
$F_D(P)$ is the average runtime we can always guarantee with
a good deterministic algorithm, provided the distribution of inputs
is known.

\medskip
\noindent{\em Yao's Result}
\medskip

Yao~(1977) claims that $F_D(P) = F_R(P)$ follows from the minimax theorem
of John von Neumann (1928).
The minimax theorem is familiar from the theory of two-person zero-sum games.

\medskip
\pagebreak[4]
\noindent{\em So What~?}
\medskip

Yao's result should not discourage the use of probabilistic
algorithms~-- we have already given several examples where they out-perform
known deterministic algorithms, and there are many similar examples.

\medskip
Yao's computational model is very restrictive.
Because $n$ is fixed, table lookup is permitted, and the
maximum complexity of any problem is $O(n^2)$.

\medskip
\noindent{\em Adleman and Gill's result}
\medskip

Less restrictive models have been considered by
Adleman and Gill.	%
Without going into details of the definitions, they prove:
\medskip

\noindent{\em Theorem:} If a Boolean function has a randomised,
polynomial-sized circuit family, then it has a deterministic,
polynomial-sized circuit family.
\medskip

There are two problems with this result:
\begin{itemize}
\item The deterministic circuit may be larger (by a factor of about $n$, the
number of variables) than the original circuit.
\item The transformation is not ``uniform''~-- it can not be computed
in polynomial time by a Turing machine.
The proof of the theorem is by a counting
argument applied to a matrix with $2^n$ rows,
so it is not constructive in a practical sense.
\end{itemize}

\section{The Class $RP$}	%

We can formalise the notion of a probabilistic algorithm and define a
class $RP$ of languages $L$ such that $x \in L$ is {\em accepted}
by a probabilistic algorithm in polynomial time
with probability $p \ge 1/2$ say\footnote{Any fixed value in $(0,1)$ can be used
in the definition.}, %
but $x \notin L$ is never accepted.
Clearly
\[ P \subseteq RP \subseteq NP, \]
where $P$ and $NP$ are the well-known classes of problems which
are accepted in polynomial time by deterministic and nondeterministic
(respectively) algorithms.
\medskip

It is plausible that
\[ P \subset RP \subset NP, \]
but this would imply that $P \ne NP$, so it is a difficult question.

\section{Perfect Parties}	%

B.~McKay (ANU) and S.~Radziszowski are interested
in the size of the largest ``perfect party''.
Because people at parties tend to cluster in groups of five,
we consider a party to be {\em imperfect} if there are five
people who are mutual acquaintances, or five who are mutual
strangers.
A {\em perfect} party is one which is not imperfect.
\medskip

McKay {\em et al} have performed a probabilistic computation which shows
that, with high probability, the largest perfect party has 42
people.

\medskip
\noindent{\em Ramsey Numbers}
\medskip

$R(s,t)$ is the smallest $n$ such that each graph on $n$ or more
vertices has a clique of size $s$ {\em or} an independent set of size $t$.
\medskip

Examples: $R(3,3) = 6$, $R(4,4) = 18$, $R(4,5) = 25$,
and $43 \le R(5,5) \le 49$.
See~\cite{McK91,McK94}.
\medskip

Perfect party organisers would like
to know $R(5,5) - 1$.

\medskip
\noindent{\em The Computation}
\medskip

A $(5,5,n)$-graph is a graph with $n$ vertices, no clique of size~5,
and no independent set of size~5.
There are 328 known $(5,5,42)$-graphs, not counting complements as different.
McKay {\em et~al} generated 5812 $(5,5,42)$-graphs using simulated annealing,
starting at random graphs. All 5812 turned out to be known.
\medskip

If there were any more $(5,5,42)$-graphs, and if the simulated annealing
process is about
equally likely to find any $(5,5,42)$-graph\footnote{There is no
obvious way to prove this, so the probability estimate is not rigorous.},
then another such graph would have been found
with probability greater than
\[ 0.99999998 \]

Thus, there is convincing evidence that all $(5,5,42)$-graphs are known.
None of these graphs can be extended to $(5,5,43)$-graphs.
Thus, it is very unlikely that such a graph exists, and it is very likely that
\[ R(5,5)-1 = 42 \]

\noindent{\em A Rigorous Proof ?}
\medskip

A rigorous proof that $R(5,5)-1 = 42$ would take thousands of years of
computer time\footnote{Based on the fact that it took seven years of
Sparcstation time to show that $R(4,5) = 25$.},
so the probabilistic argument is the best that is feasible at present,
unless we can get time on a computer as fast as
{\em Deep~Thought}~\cite{Adams}.

\section{Omissions}

We did not have time to mention applications of randomness to
serial or parallel algorithms for:
\begin{itemize}
\item sorting and selection,	%
\item computer security,	%
\item cryptography,		%
\item computational geometry,	%
\item load-balancing,		%
\item collision avoidance,	%
\item online algorithms,	%
\item optimisation,		%
\item numerical integration,	%
\item graphics and virtual reality,	%
\item avoiding degeneracy,	%
\item approximation algorithms for NP-hard problems,	%
\end{itemize}
\noindent and many other problems.
References to most of these applications are given in the bibliography
below (see for example~\cite{Mot95,Rag90}).

\medskip
\noindent{\em Another Omission}
\medskip

We did not discuss algorithms for
generating pseudo-random numbers~--
that would require another
talk\footnote{``Anyone who considers arithmetical
methods of producing random digits
is, of course, in a state of sin.''
(John von Neumann, 1951).}.

\section{Conclusion}

\begin{itemize}
\item Probabilistic algorithms are useful.
\item They are often simpler and use less space than deterministic algorithms.
\item They can also be faster, if
      we are willing to live with a minute probability of error.
\end{itemize}

\noindent{\em Some Open Problems}
\medskip

\begin{itemize}
\item Give good lower bounds for the complexity of probabilistic algorithms
(with and without error) for interesting problems.
\item Show how to generate independent random samples from interesting
structures (e.g.~finite groups defined by relations,
various classes of graphs, $\ldots$) to provide a foundation for
probabilistic algorithms on these structures.
\item Consider the effect of using pseudo-random numbers instead of
genuinely random numbers.
\item Extend Yao's results to a more realistic model of computation.
\item Give a uniform variant of the Adleman-Gill theorem.
\item Show that $P \ne RP$ (hard).
\end{itemize}

\subsection*{Acknowledgements}

Thanks to Michael Rabin for interesting me in the topic,
to George Havas for information on minimal perfect hashing,
to Brendan McKay for permission to mention his unpublished work on $R(5,5)$,
to Prabhakar Raghavan for postscript versions of his notes~\cite{Rag90}
and book~\cite{Mot95},
to John Slaney for motivating the definition of a perfect party,
and to Antonios Symvonis for
inviting me to speak at {\em Theory Day} (UNSW, April 1994)
and thus motivating me to prepare this material.
\medskip

The following bibliography is intended to help the reader follow up the
topics outlined above.  Much more extensive bibliographies can be found
in~\cite{Gup94,Mot95,Rag90}.

\end{document}